%% file: conference_101719.tex
\documentclass[a4paper,conference]{IEEEtran} 
\usepackage{cite}
\usepackage{amsmath,amssymb,amsfonts}
\usepackage{algorithmic}
\usepackage[ruled]{algorithm2e}
\usepackage{graphicx} 
\usepackage{textcomp}
\usepackage{xcolor}

\definecolor{modifica}{RGB}{0,0,0}
\graphicspath{{./images/}}

\usepackage{cite}
\usepackage{amsmath,amssymb,amsfonts}
\usepackage{algorithmic}
\usepackage{graphicx}
\usepackage{textcomp}
\usepackage{xcolor}
\def\BibTeX{{\rm B\kern-.05em{\sc i\kern-.025em b}\kern-.08em
    T\kern-.1667em\lower.7ex\hbox{E}\kern-.125emX}}

\input{defs}
\include{macros3}

\def\BibTeX{{\rm B\kern-.05em{\sc i\kern-.025em b}\kern-.08em
    T\kern-.1667em\lower.7ex\hbox{E}\kern-.125emX}}
\begin{document}
\newcommand{\JC}[1]{}
\newcommand{\DG}[1]{}
\newcommand{\AdB}[1]{}

\title{Are \emph{Multiple} Cross-Correlation Identities better than just \emph{Two}? Improving the Estimate of Time Differences-of-Arrivals from Blind Audio Signals
%{\footnotesize \textsuperscript{*}Note: Sub-titles are not captured in Xplore and
%should not be used}
%\thanks{Identify applicable funding agency here. If none, delete this.}
}
\newcommand{\ea}{\textit{et al.}~}

\author{\IEEEauthorblockN {Danilo Greco$^{\star,\dagger}$}
\IEEEauthorblockA{\textit{$^{\star}$Università degli Studi di Genova} \\
\textit{(DITEN)}\\
Via All'Opera Pia, 11a \\16145 Genova, Italy \\
danilo.greco@\{edu.unige.it, iit.it\}}
\and
\IEEEauthorblockN{Jacopo Cavazza$^{\dagger,\ddagger}$}
\IEEEauthorblockA{$^{\dagger}$\textit{Pattern Analysis and Computer Vision (PAVIS)} \\\textit{Istituto Italiano di Tecnologia (IIT)}\\ Via Enrico Melen, 83 \\  16152 Genova, Italy \\ jacopo.cavazza@iit.it}
\and
\IEEEauthorblockN{Alessio Del Bue$^{\dagger,\ddagger}$}
\IEEEauthorblockA{$^{\ddagger}$\textit{Visual Geometry and Modelling (VGM)}\\\textit{Istituto Italiano di Tecnologia (IIT)}\\
Via Enrico Melen, 83 \\ 16152 Genova, Italy \\
alessio.delbue@iit.it}
\and

%\author{\IEEEauthorblockN{1\textsuperscript{st} Danilo Greco$^{\ast \dagger}$}
%\IEEEauthorblockA{\textit{$^{\ast}$DITEN} \\
%\textit{Università degli Studi di Genova}\\
%Via All'Opera Pia, 11a \\16145 Genova, Italy \\
%danilo.greco@edu.unige.it\\$^{\dagger}$\textit{PAVIS}\\\textit{Istituto Italiano di Tecnologia (IIT)}\\Via Enrico Melen, 83\\  16152 Genova, Italy\\danilo.greco@iit.it}
%\and
%\IEEEauthorblockN{2\textsuperscript{nd} Jacopo Cavazza}
%\IEEEauthorblockA{\textit{PAVIS} \\\textit{Istituto Italiano di Tecnologia (IIT)}\\ Via Enrico Melen, 83 \\  16152 Genova, Italy \\ jacopo.cavazza@iit.it}
%\and
%\IEEEauthorblockN{3\textsuperscript{rd} Alessio Del Bue}
%\IEEEauthorblockA{\textit{PAVIS, VGM}\\\textit{Istituto %Italiano di Tecnologia (IIT)}\\
%Via Enrico Melen, 83 \\ 16152 Genova, Italy \\
%alessio.delbue@iit.it}
%\and

%\IEEEauthorblockN{4\textsuperscript{th} Given Name Surname}
%\IEEEauthorblockA{\textit{dept. name of organization (of Aff.)} \\
%\textit{name of organization (of Aff.)}\\
%City, Country \\
%email address or ORCID}
%\and
%\IEEEauthorblockN{5\textsuperscript{th} Given Name Surname}
%\IEEEauthorblockA{\textit{dept. name of organization (of Aff.)} \\
%\textit{name of organization (of Aff.)}\\
%City, Country \\
%email address or ORCID}
%\and
%\IEEEauthorblockN{6\textsuperscript{th} Given Name Surname}
%\IEEEauthorblockA{\textit{dept. name of organization (of Aff.)} \\
%\textit{name of organization (of Aff.)}\\
%City, Country \\
%email address or ORCID}
}

\maketitle

\begin{abstract}
\textcolor{modifica}{Given an unknown audio source, the estimation of time differences-of-arrivals (TDOAs) can be efficiently and robustly solved using blind channel identification and exploiting the cross-correlation identity (CCI). % which makes acquisition independent from the ordering of the microphones. 
Prior ``blind'' works have improved the estimate of TDOAs by means of different algorithmic solutions and optimization strategies, while always sticking to the case $N = 2$ microphones. But what if we can obtain a direct improvement in performance by \emph{just} increasing $N$?\\
In this paper we try to investigate this direction, showing that, despite the arguable simplicity, this is capable of (sharply) improving upon state-of-the-art blind channel identification methods based on CCI, without modifying the computational pipeline. Inspired by our results, we seek to warm up the community and the practitioners by paving the way (with two concrete, yet preliminary, examples) towards joint approaches in which advances in the optimization are combined with an increased number of microphones, in order to achieve further improvements.}

\end{abstract}

\begin{IEEEkeywords}
Acoustic Impulse Response, Blind Channel Identification, Incremental and Ensembling Approaches, TDOA Estimation
\end{IEEEkeywords}

\section{Introduction}

\begin{figure}[t!]
\centering
\includegraphics[width=\columnwidth]{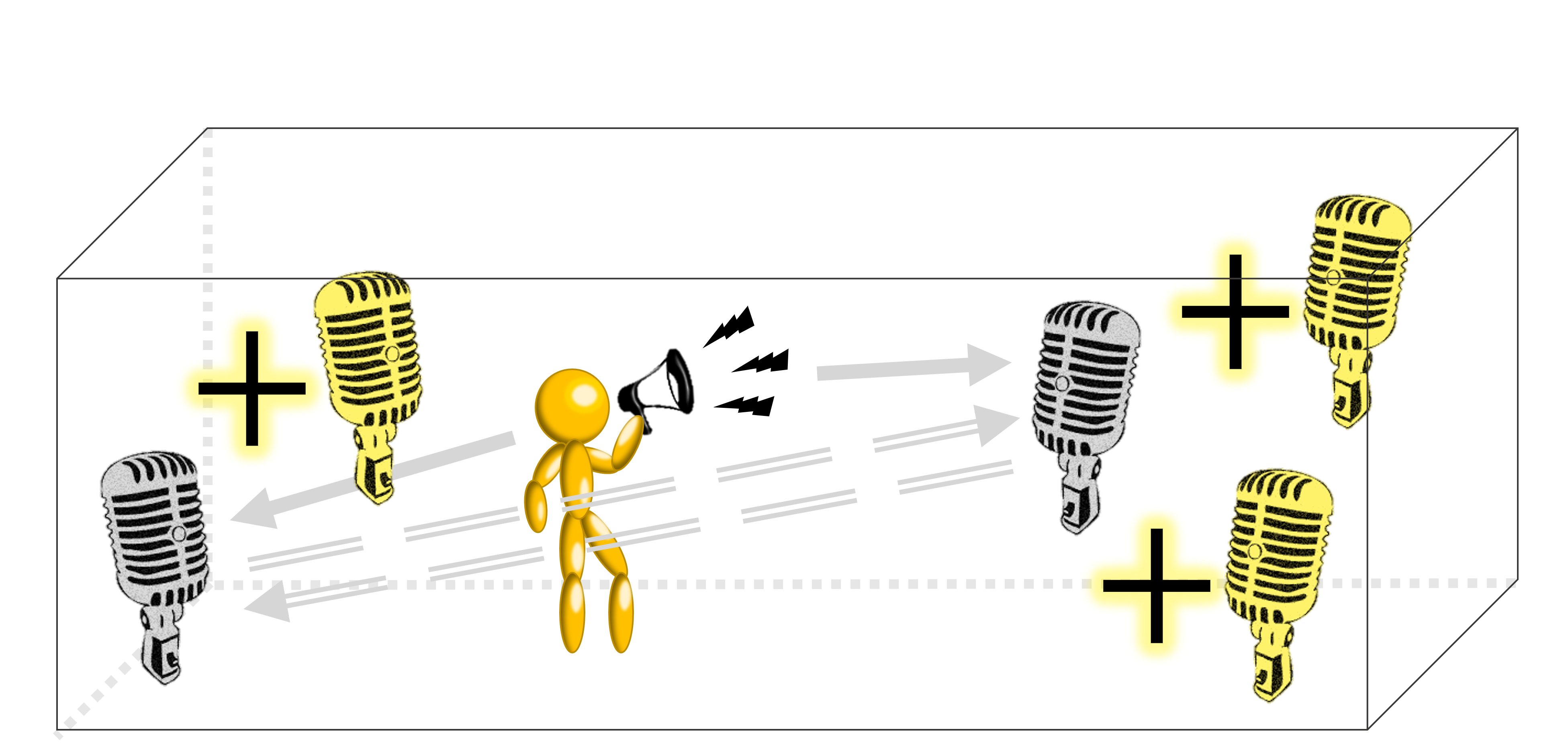}
\caption{We are given an unknown sound-emitting source, where in the actual applicative scenario that we encompass, we have no prior knowledge about the sound source and can be therefore arbitrary. We are interesting in (robustly) inferring TDOAs in an (unknown as well) environment given a pool of microphones, using the the following principle. Given the pair of grey microphone, the audio that each of them acquires from the source (solid arrow) must ``agree'' with the other. That is, if any of the two mic could ``hear'' the other, the registered signal has to be the very same (dashed arrows). This is called \emph{cross-correlation identity} and it was empirically studied in the case $N =2$, only. In this paper we answer to what happens then if $N > 2$? Can we improve in robustness and/or accuracy in the estimate, for instance, by adding the yellow microphones?}
\label{fig:CCI}
\end{figure}

Sound source localisation applications can be tackled by inferring the time-difference-of-arrivals (TDOAs) between a sound-emitting source and a set of microphones. Among the referred applications, one can surely list room-aware sound reproduction \cite{b1}, room geometry's estimation  \cite{b2,b3,b4,b5,b6}, speech enhancement \cite{b7,b21} and de-reverberation \cite{b8,b9,b10}. \textcolor{modifica}{Despite a broad spectrum of prior works estimate TDOAs from an known audio source \cite{KleinVo,Cavallaro,LasVegas}, even when the signal emitted from the acoustic source is \emph{unknown}, TDOAs can be inferred by comparing the signals received at two (or more) spatially separated microphones \cite{b10,b11,b14,b17,b18} using the notion of \textbf{cross-corrlation identity} (CCI) - see Fig. \ref{fig:CCI}. This is the key theoretical tool, not only, to make the ordering of microphones irrelevant during the acquisition stage, but also to solve the problem as \emph{blind} channel identification \cite{b10,b11,b14,b17,b18}, robustly and reliably  inferring TDOAs from an unknown audio source (see Sec. \ref{sec:probl-stat-nd-relwork}).}
%Interestingly, even when the signal emitted from the acoustic source is unknown, TDOAs can be inferred by comparing the signals received at two (or more) spatially separated microphones.This was shown to be feasible by a family of computational methods \cite{b10,b11,b14,b17,b18} which leverage the notion of \emph{cross-correlation identity} (CCI) - see Fig. \ref{fig:CCI}. That is, when handling an arbitrary pair of microphones, when the first one is behaving as a ``surrogate'' sound-emitting source, the second one has to register the very same signal as if the two microphones would have been swapped (for a more formal introduction on cross-correlation identity, the reader can refer to Sec. \ref{sec:probl-stat-nd-relwork}).

However, when dealing with natural environments, such ``mutual agreement'' between microphones can be tampered by a variety of audio ambiguities such as ambient noise. Furthermore, each observed signal may contain multiple distorted or delayed replicas of the emitting source due to reflections or generic boundary effects related to the (closed) environment. Thus, robustly estimating TDOAs is surely a challenging problem and CCI-based approaches cast it as single-input/multi-output blind channel identification \cite{b10,b11,b14,b17,b18}. Such methods promotes robustness in the estimate from the methodological standpoint: using either energy-based regularization \cite{b11}, sparsity \cite{b10,b17,b18} or positivity constraints \cite{b17}, while also pre-conditioning the solution space \cite{b10}.

In this paper, we posit that there is a much easier practical strategy to ensure robustness while inferring TDOAs: \emph{the possibility of exploiting a larger pool of microphones}. In fact, it is surprising to observe that, in prior state-of-the-art methods based on CCI, experimental evidences are provided for the case $N = 2$ microphones only \cite{b10,b11,b14,b17,b18}. Despite such a number is the bare minimum to solve the problem, it remains elusive whether $N > 2$ can, \emph{by itself}, boost the estimate of TDOAs in accuracy/robustness, without requiring any changes in the computational pipeline. In fact, since all methods \cite{b10,b11,b14,b17,b18} can \emph{theoretically} accommodate for $N > 2$, why not test them in such a regime?

The purpose of this work is to answer this question and back up the investigation of state-of-the-art methods based on CCI \cite{b10,b11,b14,b17,b18} in handling the case $N > 2$. Our goal is to understand whether an increase in the number of microphones will translate into an improved TDOAs estimate. 

{\bf Our contributions.} Among all state-of-the-art methods based on CCI \cite{b10,b11,b14,b17,b18}, we consider the most effective one: IL1C \cite{b10}. \textcolor{modifica}{Despite, in fact, recent advances were essentially devoted in estimating TDOAs given a known audio source   in how to exploit the TDOAs \cite{KleinVo,Cavallaro,LasVegas}, the problem of achieving the very same task while being blindly unaware of which audio source was deployed can be still efficiently and effectively solved using methods such as \cite{b10,b11,b14,b17,b18} out of which IL1C \cite{b10} is the best in terms of robustness and efficacy. IL1C} infers TDOAs by solving a stack of convex problems through a weighted sparsity promoting ($\ell^1$) constraint, leveraging the non-negativity of the Acoustic Impulse Response (AIR), from which TDOAs are easily estimated using peak finding \cite{b10}. To guarantee robustness while inferring TDOAs, in addition to sparsity, IL1C \cite{b10} takes advantage of a pre-conditioning mechanism to better initialize the AIRs using a data-driven initialization. 

We setup a broad experimental validation, measuring the performance of IL1C on a variety of audio signals, going well beyond the experimental evidences provided in \cite{b10}. That is, on the one side, we test the effectiveness of this method on many more audio signals: synthetic (pink and white) noise and a list of natural audio sources (two different plastic rustles -- obtained from either scraping a bag or compacting a bottle before thrashing, adult male voice, dog barking, stapler and hand-clapping). On the other side, differently to \cite{b10}, we do not only consider the case $N =2$, but we also consider a bigger number of microphones $N=3, 4, 5, 10$ motivated by encompassing the scenario of multiple microphones.
%two different necessities. On the one hand, we would like to pave to way towards a more structured acquisition setup in which multiple devices are present. On the other hand, we must still keep and eye on the real feasibility of the referred setup: therefore, we deem no interesting in going beyond $N = 10$.

As our experimental evidences show, we stably register improvements in either the robustness (towards outliers) or the accuracy in retrieving the peaks of the AIRs. We evaluate on that by exploiting two well known performance metrics as in prior work \cite{b10,b11,b14,b17,b18}, and, although there are (sound-specific) cases in which one of the two indicators show a damaged performance, still the other one shows improvements. In fact, we can demonstrate that, across the wide number of different audio sources that we consider, the general trend is that, while averaging the absolute improvement across different choices for $N =3, 4, 5$ or $6$ over the baseline case $N =2$, we score positive signed improvements (see Table \ref{tab:improvementsIL1C}) which seems not to be effected on whether the source is emitting synthetic or natural sounds. At the same time, we register a very positive trend if we are enriched by an oracle knowledge of the optimal number of microphones that have to be arranged before the acquisition stage. In such a case, we \emph{always} register positive improvements over the baseline $N = 2$, which are, in the worst case, by $+3\%$, while achieving more than $+28\%$ as well.

Inspired by our evidences, in Section \ref{sez:future_perspectives}, we attempt to warm up future research directions towards optimization approaches which explicitly account for the case $N > 2$. Although proposing a new paradigm which falls inside this new family of methods is out of scope for us, we still deem interesting to inform practitioners about the effect of two straightforward modifications of IL1C \cite{b10}, using either an incremental pre-conditioning or an ensemble strategy - see Section \ref{sez:future_perspectives}. Regardless of the scores results (in which the ensemble strategy is better than the incremental pre-conditioning, while also improving the baseline IL1C method \cite{b10}), we deem our effort to be effective in stimulating the research towards methods which %try to improve upon the estimate of TDOAs by jointly advancing the optimization approaches while also 
explicitly account for the case $N > 2$ when dealing with an unknown audio source.

%{\bf Outline of the Paper.} The rest of the paper is outlined as follows. In Section \ref{sec:probl-stat-nd-relwork} we describe the general problem and describe the related work. In Section \ref{sec:prop_method}, we describe the proposed methods , while in Section \ref{sec:experiments} we present the results obtained. In Section \ref{sec:conclusions}, we draw our conclusions.

\section{Problem Statement \& Related Work}\label{sec:probl-stat-nd-relwork}

Let us formalize the problem of inferring TDOAs, so that we can easily refer to prior related works. Let us consider a given enviroment (e.g., a room) of unknown geometry in which an audio source emits together with $N$ microphones: the task is to reconstruct TDOAs. 

Let $\vec{h}_n$ represent the AIR (Acoustic Impulse Response) from a fixed audio source and the $n$-th microphone, $n = 1,\dots,N$. The signal $\vec{h}_n$ is sampled into a fixed number of temporal bins $\vec{h}_n(k)$. The signal $y_n(k)$ received at microphone $n$ can be written as the discrete convolution between the transmitted signal $x(k)$ and the $n$-th AIR:
\begin{equation}
y_n(k) =  \vec{h}_n(k) \ast x(k)+\nu_n(k), \hspace{0.5 cm} n = 1,\ldots,N
\label{eq:conv}
\end{equation}
where $\nu_n(k)$ is an additive noise term. The ultimate goal of the problem is leveraging the measurements $y_n(k)$ to recover the AIRs $\vec{h}_n(k)$ without knowing the transmitted signal $x(k)$. 

\vspace{5 pt}

\noindent \emph{Cross-correlation identity}. When multiple microphones are recording the same audio source, the acquisition should be independent of the order of the microphones according to the following constraint:
\begin{equation}\label{eq:cri}
    \vec{h}_m(k)~\ast~\vec{h}_n(k)~\ast~x(k) = \vec{h}_n(k)~\ast~\vec{h}_m(k)~\ast~x(k),
\end{equation}
for every pairs of microphones $m$ and $n$. In turn, using eq. \eqref{eq:conv}, we rewrite eq. \eqref{eq:cri} as $\vec{h}_m(k)~\ast~y_n(k) = \vec{h}_n(k)~\ast~y_m(k)$. Hence, by using the well-known fact that the convolutional operator $\ast$ is linear, we obtain
\begin{equation}\label{eq:cri-mat}
    \mat{Y}_n \vec{h}_m = \mat{Y}_m \vec{h}_n, \hspace{1cm} m,n = 1,\dots,N
\end{equation}
where $\vec{h}_n$ is the column vector which stacks the AIRs $\vec{h}_n(k)$ by columns, while $\mat{Y}_n$ is the diagonal-constant matrix with first row and column given by $[y_n(k - K + 1), y_n(k - K), \ldots , y_n(k - K - L + 2)]$ and $[y_n(k - K + 1), y_n(k - K +2), \ldots , y_n(k), 0, \ldots , 0]^\top$ respectively, with $K$ and $L$ being the signal length and channel length. %It allows to rewrite a convolution as a linear operation \cite{b12}.
%\JC{wrong in my opinion: first row and col of $\mat{Y}_n$ cannot depend upon $k$}
%\JC{$\bullet$ 
%- What is the difference between $K$ and $L$? Should be defined since the very beginning of this Section.

In order to solve for \eqref{eq:cri-mat}, a number of prior approaches %\JC{add other refs if possible}%
have took advantage of regularization \cite{b11,b10}. For instance, Tong \ea \cite{b11} have framed the problem of TDOAs estimation as the following regularized Least Squares fitting
\begin{equation}\label{eq:tong94}
\underset{\vec{h}_1,\dots,\vec{h}_N}{min}\sum_{m\neq n}\|\mat{Y}_n\vec{h}_m-\mat{Y}_m\vec{h}_n\|_2^2 \hspace{0.4 cm} {\rm s.t.} \sum_i \| \vec{h}_i \|_2^2 = 1,
\end{equation}
to ensure robustness by means of regularization. Clearly, adding a regularization term is fundamental to avoid the optimization to converging towards the trivial solution $\vec{h}_n = 0$ for every $n = 1,\dots,N.$ Remarkably, the real problem is choosing a proper regularization term. 

In fact, when using $\ell^2$ regularization - as in eq. \eqref{eq:tong94}, the solution can be computed in closed-form by means of eigenvalue decomposition \cite{b11}. Unfortunately, $L^2$ regularization neglects some crucial physical properties of the expected solution - such as non-negativity \cite{b13,b14}.

Additionally, requiring $\sum_i \|\vec{h}_i\|^2_2 = 1$ as in \eqref{eq:tong94} makes the AIRs to be co-prime\cite{b15} %\footnote{\DG{Jacopo forse è meglio che questa nota la fai tu che sei matematico dentro..} }
and constraint each of them to have a fixed norm - each of such requirements are likely to introduce numerical instabilities and artifacts during the optimization process. As a remedy for this, sparsity priors have been successfully applied to a broad spectrum of prior work in TDOAs estimation~\cite{b1,b2,b3,b4,b5,b6} \cite{b15}, while also encompassing speech enhancement \cite{b16} and de-revereberation \cite{b8}. Therefore, as to impose sparsity in the reconstructed $\vec{h}_n$, replacing the $L^2$  regularization in eq. \eqref{eq:cri-mat} with a $L^1$ counterpart \cite{b17,b18,b10} seems an appealing solution. Precisely, in \cite{b18} a $L^1$-norm penalty was added to eq. \eqref{eq:tong94}, yielding
\begin{equation}\label{eq:kow13}
\underset{\vec{h}_1,..,\vec{h}_N}{min}\sum_{m\neq n}\|\mat{Y}_n\vec{h}_m-\mat{Y}_m\vec{h}_n\|_2^2 \hspace{0.1cm} s.t. \begin{cases} \sum_i\|\vec{h}_i\|_2^2 = 1, \\ \sum_i \|\vec{h}_i\|_1 < \varepsilon. \end{cases}.
\end{equation}
Unfortunately, a quadratic optimization subject to mixed quadratic and linear constraints do not preserve the convexity of \eqref{eq:tong94}. Hence, the method as in \eqref{eq:kow13} is prone to local solutions.

To cope with this issue, we can relax eq. \eqref{eq:cri-mat} into
\begin{equation}\label{eq:Lin07}
\underset{\vec{h}_1,..,\vec{h}_N}{min}\sum_{m\neq n}\|\mat{Y}_n\vec{h}_m-\mat{Y}_m\vec{h}_n\|_2^2 \hspace{0.1cm} s.t. \begin{cases} |\vec{h}_1(a)| = 1, \\ \sum_i \|\vec{h}_i\|_1 < \varepsilon. \end{cases},
\end{equation} where the fixed index $a$ is an anchor constraint \cite{b17} which makes the optimization in eq. \eqref{eq:Lin07} convex and more robust towards spectrum holes of $x(k)$ if compared to eq. \eqref{eq:tong94}. 

However, the anchor constraints $|\vec{h}_1(a)| = 1$ together with $\sum_i \| \vec{h}_i \|_1 < \varepsilon$ penalizes all the peaks intensities but one, often leading to peak cancellations in noisy conditions.
The approach of \cite{b17} has been modified in \cite{b14} adding an ancillary non-negativity constraint on the AIRs
\begin{equation}\label{eq:Lin07NNeg}
\underset{\vec{h}_1,..,\vec{h}_N}{min}\sum_{m\neq n} \hspace{-0.1cm} \|\mat{Y}_n\vec{h}_m-\mat{Y}_m\vec{h}_n\|_2^2 \hspace{0.1cm} s.t. \begin{cases} |\vec{h}_1(a)| = 1, \\ \sum_i \|\vec{h}_i\|_1 < \varepsilon \\ 
\vec{h}_1,..,\vec{h}_N\geq0
. \end{cases},
\end{equation}
where, for each $n$, $\vec{h}_n\geq0$ means $\vec{h}_n(k)\geq0$ for each $k$. Non-negativity yields increased robustness against noise by further regularizing the problem  \cite{b19,b20}, but it is arguably limited in addressing the limitations of the anchor constraints.

To directly tackle the latter problem, Crocco \ea \cite{b10} replaced the anchor constrained $|\vec{h}_1(a)| = 1$ by means of the introduction of a slack variables $\vec{p}_1,\dots,\vec{p}_N$ such that
\begin{equation}\label{eq:Crocco16}
\underset{\vec{h}_1,..,\vec{h}_N}{min}\sum_{m\neq n} \hspace{-0.1cm} \|\mat{Y}_n\vec{h}_m-\mat{Y}_m\vec{h}_n\|_2^2 \hspace{0.1cm} s.t. \begin{cases} \vec{p}_n^{\hspace{0.5mm}\top}{\vec{h}}_n = 1, \\ \sum_i \|\vec{h}_i\|_1 < \varepsilon \\ 
\vec{h}_1,..,\vec{h}_N\geq0
. \end{cases}
\end{equation}
In this way, all the components of the AIRs are equally taken into account without privileging the $a$-th of the $\vec{h}_1$. At the same time, differently from eqs. \eqref{eq:kow13}, \eqref{eq:Lin07}, the constraints as in eq. \eqref{eq:Crocco16} are differentiable, since $\vec{h}_1,..,\vec{h}_N\geq0$ implies $\sum_i \| \vec{h}_i \|_1 = \sum_i \sum_a \vec{h}_i(a).$ 
The optimization problem as in eq. \eqref{eq:Crocco16} is convex with respect to $\vec{h}_n$ while fixing the slack variables $\vec{p}_n$ and vice-versa. Inspired by this consideration, Crocco \ea \cite{b10} proposed an alternated iterative scheme in which, $\vec{p}_n$ are firstly initialised as the AIRs computed using Tong \ea method's \cite{b11}, while cycling between: 1) optimizing for $\vec{h}_1,..,\vec{h}_N$ in \eqref{eq:Crocco16} given $\vec{p}_1,..,\vec{p}_N$ and 2) use the newly computed AIRs to update $\vec{p}_n$ for every $n$. As discussed in \cite{b10}, although the proposed initialization introduces a distortion in the amplitude of the AIRs, then the iterative procedure is able to compensate.
More crucially, initializing $\vec{p}_n$ at the first iteration by using \cite{b14} makes the slack variable sparse. Therefore, the first two constraints as in eq. \eqref{eq:Crocco16} make the computed AIRs sparse again. Such property is preserved during optimization because of the updating scheme in which slack variables at a given iteration are selected as the solution of eq. \eqref{eq:Crocco16} as in the prior iteration.

\vspace{ 5 pt}

\noindent \underline{\emph{A sharp limitation of prior blind methods}}. None of the prior methods \cite{b10,b11,b14,b17,b18} was generalized to the case $N > 2$. Despite $N = 2$ has the appealing formal property of achieving minimality among the number of microphones necessary to solve the optimization problem, still it remains elusive from a practical standpoint whether allocating for a bigger number $N$ of microphones can effectively boost the estimate of TDOAs. And, in the likely event of this case effectively happening, are we improving upon robustness towards outliers or in accuracy as well? The scope of the present work is to answer this question.

\section{Multiple Cross-Correlation Identities}\label{sez:multiple_CCI}

In this Section, we evaluate the effect of increasing the number of microphones when tackling the problem of inferring TDOAs by means of well established notion of cross-correlation identity (CCI) \cite{b10,b11,b14,b17,b18}. In details, we focus on IL1C \cite{b10}, the best out of such class of approaches: we optimize equation \eqref{eq:Crocco16} for the case of $N = 2,3,4,5,10$. By doing so, we are capable of starting from the minimal setup from which the problem can be solved ($N = 2$): note that this is the experimental playground analysed in prior works \cite{b10,b11,b14,b17,b18}. Differently, for the sake of inspecting whether a higher number of microphones can provide an improvement in the estimate of TDOAs, we also consider the cases $N=3,4,5$ up to the $N = 10$ microphones. This range of variability in $N$ is, in our opinion, a good trade-off between having a sufficiently large number of acquisition devices, while still framing a scenario which can be still useful from the applicative standpoint. 

Let us briefly introduce the types of source signals considered in this study, as well as the reproducibility and implementation details about our evaluation protocol and the error metrics to check on performance. The results of our analysis are reported in Tables \ref{tab:all_EPP} and \ref{tab:all_PUP}, while showing relative and absolute improvements in Table \ref{tab:improvementsIL1C}. An extended discussion on our findings is reported in Section \ref{ss:discussion_IL1C_N}.

\begin{table*}
    \centering
    \caption{Average Peak Position Mismatch ($\mathcal{A}_{\rm PPM}$)) metrics for IL1C \cite{b10} when $N=2,3,4,5,10$. Synthetic source noise are denoted in italic, while bold italic refers to the natural source signal considered in this study. For each source signal considered, we provide an histogram visualization to better perceive the variability of the error metrics: the range of variability of each data bar is normalized within each different source signal. A better performance corresponds to a lower ($\mathcal{A}_{\rm PPM}$)) value or, equivalently, to a lower bar.  The value $s$ quantifies the impact of the additive Gaussian noise on the registered signal: we span the case $s=0.01$ (easier) to $s=1$ (harder), while transitioning on the intermediate cases $s=0.1,0.2$ and $s=0.5$.}
    \label{tab:all_EPP}
    \includegraphics[width = .8\textwidth]{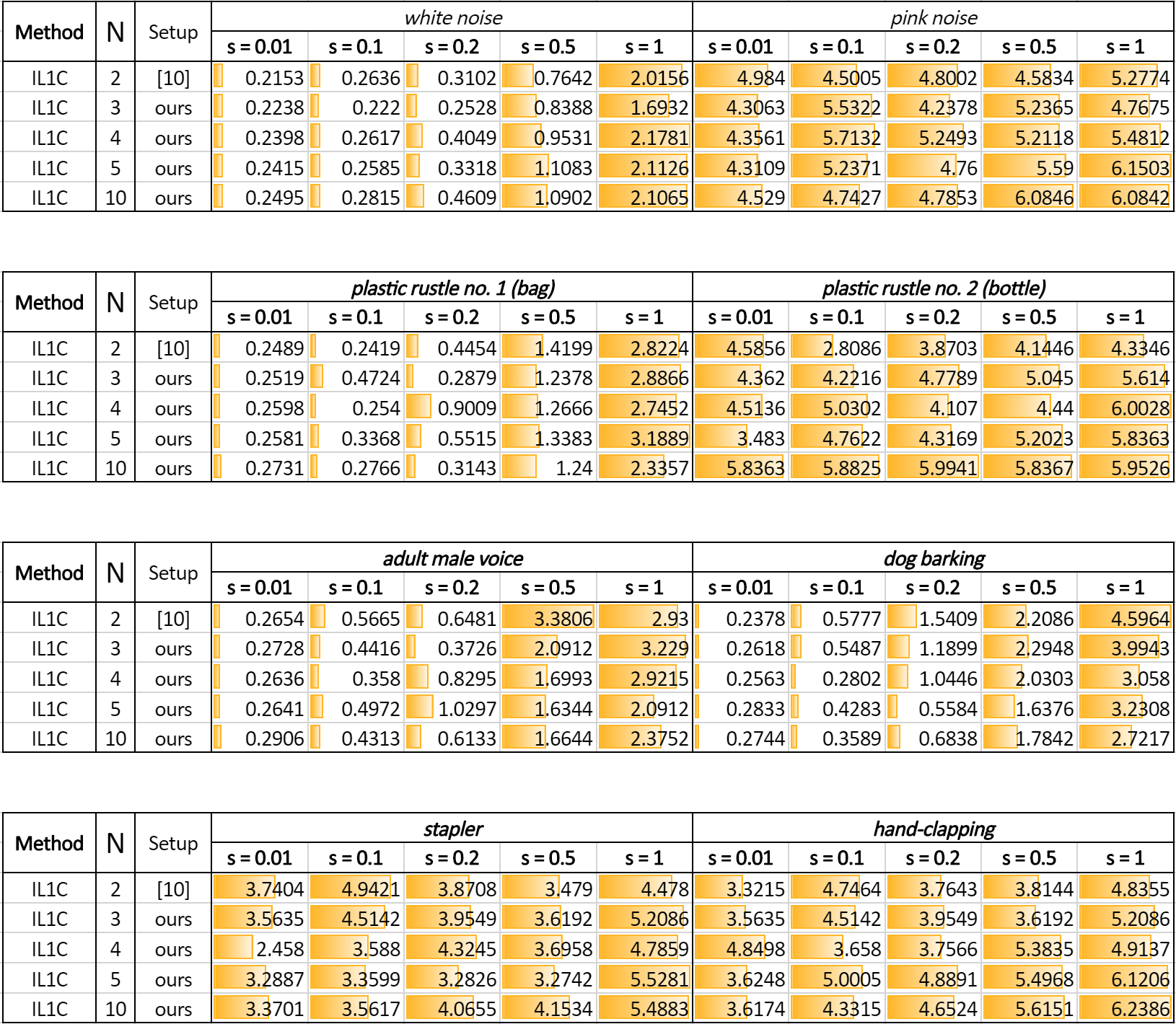}
    
\end{table*}

\begin{table*}
    \centering
    \caption{Average Percentage of Unmatched Peaks ($\mathcal{A}_{\rm PUP}$) metrics for IL1C \cite{b10} when $N=2,3,4,5,10$. Synthetic source noise are denoted in italic, while bold italic refers to the natural source signal considered in this study. For each source signal considered, we provide an histogram visualization to better perceive the variability of the error metrics: the range of variability of each data bar is normalized within each different source signal. A better performance corresponds to a lower ($\mathcal{A}_{\rm PUP}$)) value or, equivalently, to a lower bar. The value $s$ quantifies the impact of the additive Gaussian noise on the registered signal: we span the case $s=0.01$ (easier) to $s=1$ (harder), while transitioning on the intermediate cases $s=0.1,0.2$ and $s=0.5$.}
    \label{tab:all_PUP}
    \includegraphics[width = .8\textwidth]{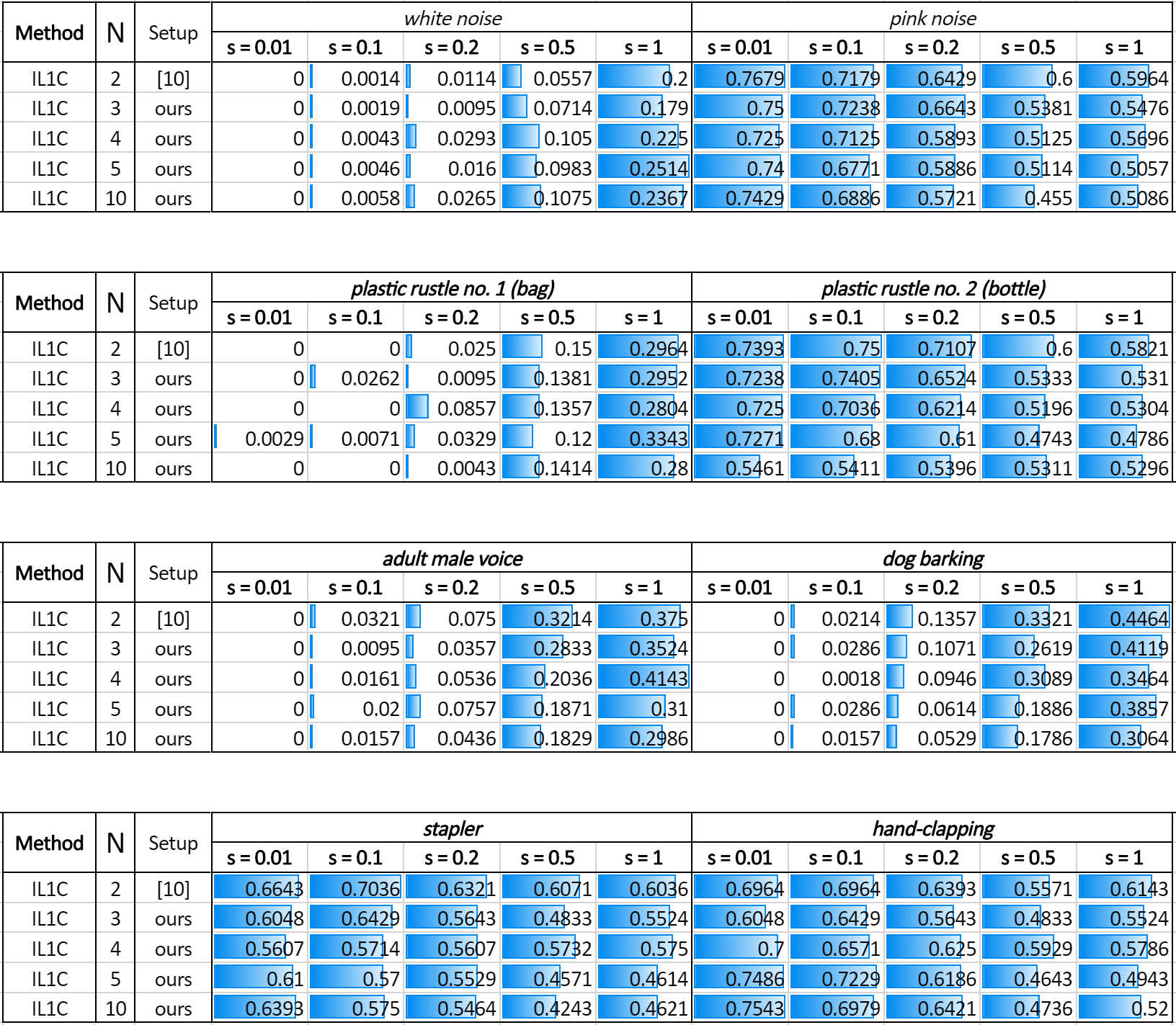}

\end{table*}

\vspace{5 pt}

\textit{The different types of source signals we considered.} %\JC{@Danilo, potresti per favore descrivere un secondo i rumori per favore, diciamo 3/5 righe per ognuno? Nei paper si deve fare di descrivere il dato dicendo dettagli su come e' stato acquisito e alcune specifiche tecniche. Qua soprattutto bisogna far capire che ci sono due gruppi di rumori sintetici e "naturali" e testiamo su tutti. Pensavo ad una cosa tipo di elenco a punti almeno alla gente e' chiaro che hai fatto test su tanti rumori diversi}
%\DG{The quantitative results over 50 Monte-Carlo simulations are reported in Table 1 for synthetic and natural, speech and non-speech, sources. The idea behind is to test the method with different signals $s(k)$ which show different behaviour in frequency spectrum (wide rather then narrow)}
\textcolor{modifica}{We considered two types of synthetic audio signals \textit{white noise} and \textit{pink noise}, which differ among each others in the considered frequencies of their spectrum (all vs. only wide ones, respectively). We also encompass a broad list of natural sounds as audio source: \textbf{\textit{plastic rustle no. 1 (bag)}},  \textbf{\textit{plastic rustle no. 2 (bottle)}},  \textbf{\textit{adult male voice}}, \textbf{\textit{dog barking}}, \textbf{\textit{stapler}} and \textbf{\textit{hand-clapping}}, all of them characterized by a narrow frequency spectrum.}

\vspace{5 pt}

\textit{Evaluation.} We run experiments by considering any of the source audio signals described in the prior paragraph located in the same environment analyzed in \cite{b10}. We model the Acoustic Impulse Response (AIR) for each microphones as seven different peaks, corresponding to one direct path source-microphones, together with six (first-order) reflections. In details, we applied the simulating image method as in \cite{b16}, using a reflection coefficient of $0.8$. We also introduce another degree of variability, by considering different Noise-to-Signal ratios ($s$). This is done by injecting additive Gaussian white noise on the output microphones according to the following specs: $0$ dB, $6$ dB , $14$ dB, $20$ dB and $40$ dB. This induces a signal-to-noise ratio $s=10^{-\mathrm{dB}/20}$ from the following inverse relationship $\mathrm{dB}=20 \log_{10}(1/s)$. %\JC{Controlla un attimo Danilo per favore. Non mi e' mai stata chiara sta roba delle soglie segnale-rumore} \DG{il segnale maggiormente affetto da rumore è quello a 0 dB che corrisponde a s=1 mentre quello meno affetto da rumore è quello a 40 dB che corrisponde a s=0.01}. In details, the value of $s$ corresponding to the previous cases are: $0.01$, $0.1$, $0.2$, $0.5$ and $1$, respectively: 
%\begin{equation}\mathrm{dB}=20 \log_{10}(1/\mathrm{s})\end{equation}
%\begin{equation}\mathrm{s}=10^{(-\mathrm{dB}/20)}\end{equation}
When running the optimization \eqref{eq:Crocco16} of IL1C \cite{b10}, we take advantage of the official code directly shared by authors, while following the same pre-processing and evaluation techniques as in the referred prior work. In addition, as done \eqref{eq:Crocco16}, we perform model selection by doing cross-validation on the threshold $\varepsilon$ which controls the sparsity-promoting constraint. 

%We run simulations %considering a rectangular room of $6 \times 5 \times 4$ m %
%with microphones and sources position randomly generated at each interaction. %; we modeled the AIR for each microphone as seven peaks in random time positions (one direct path source-microphone plus six reflections from walls). %\footnote{To avoid configurations in which the source is too close to the microphones, the $x$ coordinate ranged from $0$ m to $2$ m for the microphones and from $4$ m to $6$ m for the source.}%. 
% The AIRs were simulated according to the image method \cite{b16} assuming a reflection coefficient of $0.8$. We started using a $1$s long synthetic white noise as $x(k)$ sampled at $16$ kHz and the length of the estimated channel was of $L= 700$ samples and such a value was chosen for being an upper bound for every possible channel length, given the room geometry; then we used a real 2 s long recorded signal in a rustle due to a plastic material. Simulation were repeated varying the Signal to Noise Ratio (SNR) adding Gaussian white noise at the output microphones by setting it to the following values: $40$ dB, $20$ dB, $14$ dB $6$ dB and $0$ dB. All the parameters for all the tested methods ($\lambda$) were optimized through cross-validation as recommended in \cite{b10}.

\vspace{5 pt}

\textit{Error metrics.} Once the AIRs have been computed through \eqref{eq:Crocco16}, we apply the peak finding method of \cite{b10} and we evaluate performance by means of two standard error metrics: the Average Peak Position Mismatch ($\mathcal{A}_{\rm PPM}$) and the Average Percentage of Unmatched Peaks ($\mathcal{A}_{\rm PUP}$) \cite{b15}. To ensure statistical robustness towards the random generation of reflections using \cite{b16}, we performed $Z = 50$ random repetitions of the experiments using Monte-Carlo simulation \cite{b10}. A ground truth peak is considered to be unmatched if the closest estimated number is more than a fixed number of samples aways from it (we follow \cite{b10} in setting this value equal to 20). In formul\ae, we compute $\mathcal{A}_{\rm PPM}$ and $\mathcal{A}_{\rm PUP}$ in the following manner
\begin{align}
\mathcal{A}_{\rm PPM} & = \dfrac{1}{Z} \sum_{i=1}^{Z}\sum_{p=1}^{\bar{P}_i}\frac{|\tau_{p,i}-\widetilde{\tau}_{p,i}|}{\bar{P}_i} \label{eq:PPM} \\ 
\mathcal{A}_{\rm PUP} & = \dfrac{1}{Z} \sum_{i=1}^{Z}\frac{K-\bar{P}_i}{K} \label{eq:PUP}
\end{align} 
where $\bar{P}_i$ is the number of ground truth peaks for which a matching has been found among the estimated ones: such value is indexed over the Monte-Carlo simulations $i = 1,\dots,Z$. For every $i$ and given an arbitrary $p = 1,\dots,P_i$ $\tau_{p,i}$, in eq. \eqref{eq:PPM}, $\tau_{p,i}$ and $\widetilde{\tau}_{p,i}$ are the $p$-th ground truth peak location and its corresponding estimate, respectively. In eq. \eqref{eq:PUP}, $K$ denotes is the number of ground truth peaks of the source signal. 

By means of such metrics, we can decouple the effect of the outliers (quantified by $\mathcal{A}_{\rm PUP}$) from the  overall peak position accuracy (expressed by $\mathcal{A}_{\rm PPM}$), ultimately better evaluating on the robustness with which TDOAs are estimated. 
%Given a source signal $x(k)$ and we solve the system \eqref{sistema} in the case $N > 3$ to get the AIRs: $\vec{h}_{m_{i}}(k)$ (with either the original or the incremental IL1C) or $\vec{h}_{{m_i},AVG}$ when performing the ensemble. Leveraging the AIRs, we adopt standard error metrics to quantitatively evaluate the results: the Average Peak Position Mismatch ($A_{PPM}$) and the Average Percentage of Unmatched Peaks ($A_{PUP}$) \cite{b15}. In both cases, once a single source signal $x(k)$ has been fixed, we run $Z$ different Monte Carlo simulations, out of which the average are computed.
%These values are both calculated on the peaks corresponding to direct path and reflections. 
%We modeled the AIR for each microphone as seven peaks signal in random time positions (one direct path source-microphone plus six 1-st order reflections from the walls).
%A ground truth peak has been considered unmatched if the closest estimated peak is more than a fixed number of samples away from it (here 20, as in \cite{b10}). In formul\ae, the $A_{PPM}$ and the $A_{PUP}$  are calculated as it follows:
%The quantitative summary of the results over the $50$ Monte Carlo simulations is reported in Table 1 for synthetic, non-speech and speech sources respectively. In each box the first term is the% 

%The APUP is calculated as  
%\begin{equation}
%A_{PUP} = \sum_{i=1}^{50}\frac{K-P_i}{50K}
%\end{equation} 
%where $K$ is the number of ground truth peaks.

\section{The proposed Test-Case: a Discussion}\label{ss:discussion_IL1C_N}

\vspace{5 pt}

\emph{Performance differences across variable $s$ values.} %According to the definition 
An increasing value for $s$ will make the acquired signal noisier, so that, in Tables \ref{tab:all_EPP} and \ref{tab:all_PUP} the case $s = 0.01$ is (much) easier with respect to $s = 1$. This visually translates into errors (and histogram bars) which increase when moving from left to right in the referred error tables. A sharp increase of errors is registerd on \emph{white noise} (synthetic) and \textbf{\textit{bag plastic ruslte}}, \textbf{\textit{adult male voice}} and \textbf{\textit{dog barking}} (natural). Differently, on either pink noise (synthetic) or stapler, hand-clapping (natural), we can see that already the case $s = 0.01$ is challenging per se. We posit that a reason for that is the highly oscillatory natura of those sounds that, if compared to other cases, make them less influeced by the additive Gaussian noise (since they behave as if they were intrinsically noisy) \JC{Danilo per favore rileggi questo paragrafo bene e dimmi se possiamo trovare una spiegazione meno tirata per i capelli della mia per piacere!}  

\vspace{5 pt}

\emph{Differences between synthetic and natural sound-emitting sources.} Let us comment on whether the usage of a synthetic versus a natural source emitting sound can have an impact on the final performance. According to the experimental results reported in Tables \ref{tab:all_EPP} and \ref{tab:all_PUP}, while also inspecting the signed absolute/relative improvements of Table \ref{tab:improvementsIL1C}, we can get that there seems not to be a sharp difference in performance between these two categories. In fact, we did not register any drop/raise when swapping from white/pink noise to the other sounds considered in this work. We deem this a valuable property of the cross-correlation identity (CCI) which can naturally accommodate for a variety of applicative scenarios where the audio source is unknown. %we cannot put a constraint on the type of sound our method is asked to handle.

\vspace{5 pt}

\underline{\textit{\textbf{Does adding microphones improves upon performance?}}} We are intended in enriching this discussion with a detailed analysis on the ultimate question that our work is trying to respond. We believe that the findings of Tables \ref{tab:all_EPP} and \ref{tab:all_PUP} are plain: the honest answer to the aforementioned question we are intended to respond is neither positive nor negative, \emph{in general}. In fact, there is a quite number of cases in which the addition of microphones is not clearly beneficial, on the contrary damaging performance: for the sake of brevity, let us report the worst cases for the two metrics. That is, the cases $s=1$, $N=10$ and \textbf{\textit{hand-clapping}} for  $\mathcal{A}_{\rm PPM}$ (-1.4031 absolute improvement) and $s=1$, $N=4$ and \textbf{\textit{adult male voice}} for  $\mathcal{A}_{\rm PUP}$ (-0.0393 absolute improvement). These are definitely failure cases and, specifically, \textbf{\textit{hand-clapping}}, $s=1$ for $\mathcal{A}_{\rm PPM}$ is clearly not positive since the trend is that performance drops while $N$ increases. Albeit these cases are surely negative, let us observe that there are actually no cases where \emph{concurrently} the two metrics deteriorate. In fact, in the worst cases, only one of the two is damaged: we either loose in effectiveness on how we handle outliers or in how accurately we retrieve the peaks. But, globally the case $N > 2$ is never inferior to the baseline $N =2$ with respect to both metrics concurrently.

At the same time, let us observe that these failure cases are limited since, in the majority of the (remaining) cases, the performance is either stable (therefore adding microphones is not detrimental) or better (and thus addding microphones actually help). The fact that performance is stable when varying the number of microphones is true for the (less noisy) cases $s = 0.01$, \textit{pink noise}, for $\mathcal{A}_{\rm PPM}$; $s = 0.01$, \textbf{\textit{adult male voice}}, for both $\mathcal{A}_{\rm PPM}$ and $\mathcal{A}_{\rm PUP}$; $s = 0.01$, \textit{pink noise}, for $\mathcal{A}_{\rm PUP}$; $s = 0.01$ \textit{\textbf{dog barking}}, for $\mathcal{A}_{\rm PPM}.$ 

Finally, let us concentrate on the ideal cases, where the performance improves when $N$ raises. This happens for (the more challenging) cases such as $s=1$ \textit{\textbf{dog barking}}, for $\mathcal{A}_{\rm PPM}$; $s=0.5$ \textit{\textbf{adult male voice}}, for $\mathcal{A}_{\rm PPM}$; $s=0.1$, \textit{\textbf{stapler}}, for $\mathcal{A}_{\rm PPM}$ and $s=0.5$, \textit{\textbf{adult male voice}}, for $\mathcal{A}_{\rm PUP}$, $s=0.1$ and $s=0.2$, \textit{\textbf{plastic rustle no. 2 (bottle)}} for $\mathcal{A}_{\rm PUP}$; $s=0.2$, \textit{\textbf{dog barking}} for $\mathcal{A}_{\rm PUP}$.

Given the alternate nature of the results, when switching from one error metric to another and while varying different $s$ and $N$ values, we deem necessary to summarize the highlights of our findings in the next part of our discussion.

\vspace{5 pt}

\underline{\textit{\textbf{A summary of the improvements.}}} %The reader can refer to Table \ref{tab:improvementsIL1C} for a quick glance of the scored signed absolute/relative improvements over all the different type of source audio signals considered in this work.
In Table \ref{tab:improvementsIL1C} (bottom), we report the average absolute signed improvement $\delta_{\rm avg}$ over the two error metrics $\mathcal{A}_{\rm PPM}$ and $\mathcal{A}_{\rm PUP}$: the overall majority of the cases show a superiority of the case $N > 2$ with respect to the baseline case $N = 2$ of IL1C \cite{b10}. This is exemplified from the fact that the signed improvement is positive  ($\delta_{\rm avg} > 0$) for 5 out of 8 different audio signals, in terms of $\mathcal{A}_{\rm PPM}$, and 7 times out of 8, in terms of $\mathcal{A}_{\rm PUP}$. Despite of their sign, the absolute value of such improvements is controlled (it never exceeds $0.5$). This trend is explained from the fact that, there are high fluctuations, sometimes, between different configurations inside the case $N>2$ for an unknown audio source.

To better investigate this trend, we also consider the signed relative improvements $\Delta^O$ of the error metrics $\mathcal{A}_{\rm PPM}$ and $\mathcal{A}_{\rm PUP}$ (Table \ref{tab:improvementsIL1C}, top). In this case, we allow for an oracle selection of the best number $N$ of the microphone configuration so that we can understand what is the ``upper'' bound on the improvement that we can expect to register. The results are extremely encouraging: we \emph{always} have significant positive improvements. In the worst cases (\textit{\textbf{plastic rustle no. 2 (bottle)}}, $\mathcal{A}_{\rm PPM}$), we get a +2.8\% while, in the most favorable case (\textit{\textbf{adult male voice}}, $\mathcal{A}_{\rm PPM}$), the relative improvement sharply raises, reaching +28.4\%.

\begin{table}[t!]
    \centering
    \caption{Signed improvements for the metrics $\mathcal{A}_{\rm PPM}$ and $\mathcal{A}_{\rm PUP}$ when comparing $N > 2$ with the baseline $N=2$ using the state-of-the-art method \cite{b10}. \emph{Top}: we provide the percentage relative improvements $\Delta^O$ using the oracle selection for microphone number's configuration (reported as a superscript). \emph{Bottom}: We provide the mean absolute improvement $\delta_{\rm avg}$ across \emph{all} cases $N=3,4,5,10$ with respect to the baseline $N=2$. \emph{Top and Bottom}: We report the aforementioned statistics for the more challenging noise-to-signal ratio $s=1$.}
    \label{tab:improvementsIL1C}
    \includegraphics[width=\columnwidth]{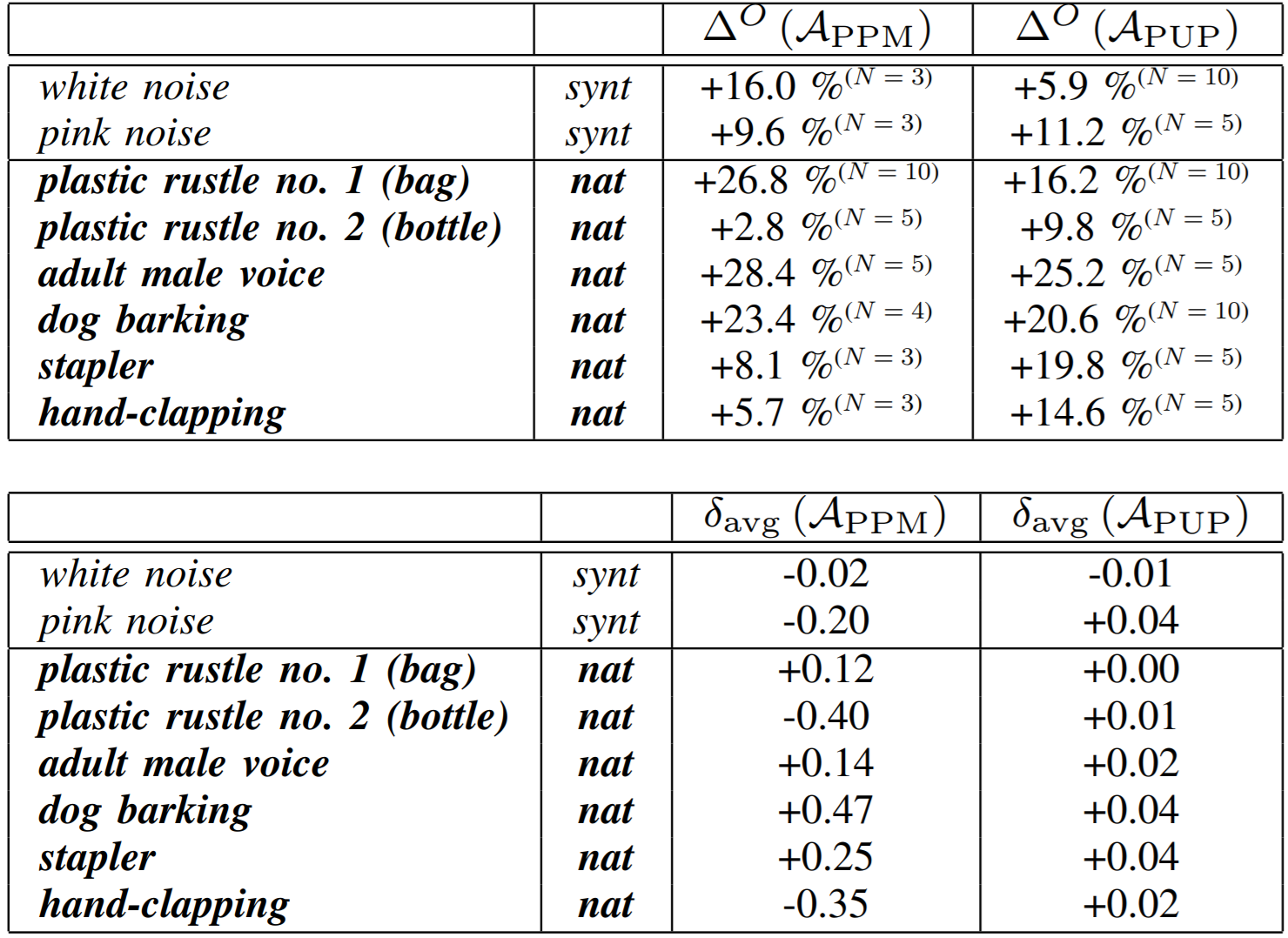}
\end{table}

\section{Future Perspectives}\label{sez:future_perspectives}

In shed of the results of our test-case (Table \ref{tab:improvementsIL1C}), we deem now reasonable for practitioners to start investigating the regime $N > 2$ (unknown source) with computational methods which take advantage of this scenario in explicit terms. Although this actual effort is beyond the scope of the present submission, we are nevertheless interested in warming up the research in this direction by considering what are, to our opinion, the easiest modification that can be applied to the state-of-the-art method IL1C \cite{b10}. In the rest of the present Section, we will present two computational variants of IL1C which are either based on an \emph{incremental pre-codintioning} or an \emph{ensemble mechanism}.

\textit{\textbf{Incremental pre-conditioning}}. Given the core contribution of pre-conditioning the solution that IL1C introduced, we can think about an \emph{incremental} preconditioning in which we gradually introduce one microphone, intertwining this operation with a fine-tuning of the AIRs. That is, we start from a pair of microphones and we optimize for it. Then, we use the solutions of IL1C for that pair to pre-condition the solution when solving for a third microphones: we the update also the AIRs for the first two microphones. The procedure iterates until the $N$-th microphones is added (so that the $N-1$ AIRs of the other microphones are fine-tuned, at least one time). 
Let us formalize the prior argument in the following pseudocode.\\
{\bf 1}. Sample two random microphones $m_1$, $m_2$.\\
{\bf 2}. Optimize eq. \eqref{eq:Crocco16}, using the \emph{standard} pre-conditioning \cite{b10}, thus obtaining the AIRs for $m_1$ $m_2$.\\
{\bf 3}. Add a third microphone $m_3$: optimize eq. \eqref{eq:Crocco16} again but now changing the preconditioning. The AIRs of $m_1$ and $m_2$ will be the ones obtained at the previous stage, while the AIR of $m_3$ will be initialized using the standard approach \cite{b10}.\\
{\bf 4}. Update the AIRs for all solved microphones.\\
{\bf 5}. Keep adding microphones, following the same procedure, until all $N$ ones are covered

\emph{Results \& Discussion}. We did not register any substantial improvement using this sequential addition, to the point that even the case $N = 2$ is superior in performance. For the sake of brevity, let us report a glance of the scored results, providing a peculiar case which is aligned with the general trend which we do not report for the sake of brevity. For \textit{white noise}, the results of incremental strategy describe above are 0.0036 ($s=0.01$), 0.0357 ($s=0.1$), 0.09 ($s=0.2$), 0.1536 ($s = 0.5$) and 0.2343 ($s=1$) for $\mathcal{A}_{\rm PUP}$ and 0.2658 ($s=0.01$), 0.5866 ($s=0.1$), 1.0023 ($s=0.2$), 1.7345 ($s=0.2$) and 	2.2391 ($s=1$) for $\mathcal{A}_{\rm PPM}$ -- all error values refer to the case with $N=4$, while averaging over $Z=50$ random extraction of the sequence with which microphones are incrementally added. We explain this lack of improvement with the fact that, despite adding microphones \emph{in a single solution} maybe beneficial, their sequential addition can be detrimental since, albeit on the one side the case $N > 2$ is providing more cues than the baseline $N =2$, the sequential addition of microphone would lead to ``over-fitting'' the AIRs of some of the microphones, ultimately damaging the final performance.

\textit{\textbf{Ensemble mechanism}}.
Let us observe that the inference stage of IL1C \cite{b10} is based on peaks finding, a method which is known to suffer when spurious peaks are present. To accommodate for that, let us take advantage of the following approach. We can split the case $N > 2$ into several $N = 2$ sub-problems, by pairing microphones into couples. We therefore create a number of playgrounds with 2 microphones only (unknown source) - so that we match the operative conditions on which IL1C \cite{b10} was originally tested. We therefore create some redundancy in the estimate of the AIRs: this is because one microphone can belong to several pairings at the same time, so there will be multiple candidate solutions for the same AIRs - two candidates, referring to two different microphones, from each artificial pairing. We solve for this redundancy by averaging out all different candidates referring to the same microphone. We deem this approach to be arguably simple, perhaps rough, but still effective in handling a well known computational issue which damages peak findings algorithm. In fact, the presence of spurious (noisy) peaks surely affect the estimate of TDOAs. We attempt to mitigate this problem by exploiting the well known smoothing and regularizing properties of averaging as our ensemble mechanism.

\emph{Results \& Discussion}. The reader can refer to Table \ref{tab:medie} for the quantitative evaluation of our ensemble strategy applied to IL1C \cite{b10} evaluated in the test-case $N = 10$. We are expecting to register a very interpretable phenomenon out of a simple strategy such as averaging multiple candidate solutions corresponding to the same AIR: we should expect to register a regularizing effect which smooths out the AIRs, removing spurious peaks due to, for instance, numerical instability. This explains the improvements achieved from our proposed ensemble mechanism versus the IL1C \cite{b10} baseline: once spurious peaks have been removed, we expect that a peak finding algorithm such that the one applied in \cite{b10} can be more effective in finalizing the estimate of TDOAs. This consistently happen in the cases $s=0.01$, $s=0.1$ (for both $\mathcal{A}_{\rm PPM}$ and $\mathcal{A}_{\rm PUP}$) and $s=0.2$ (only for $\mathcal{A}_{\rm PUP}$), while, when considering the ``more difficult'' cases $s=0.5$ and $s=1$ we do not see a sharp improvement of the ensemble method. This is probably due to the fact that the candidate solutions that are averaged are, each of them, noisier. Therefore, the averaging effect produces an excessive over-regularization which excessively smoothens the peaks, damaging the performance of the peak finding. Nevertheless, the regularizing effect of averaging can be inspirational for practitioners in exploiting a large number of microphones to better estimate TDOAs. 

\begin{table}[t!]
    \centering
    \caption{The ensemble mechanism. We the report the performance of IL1C \cite{b10} ($N=10$, \emph{white noise}) versus the ensemble mechanism in which couples of microphones are solved, first, and the aggregated by averaging across the redundancy of AIRs referring to the same microphones. We denote a better performance in bold, across different signal-to-noise values $s$.}
    \label{tab:medie}
    \begin{tabular}{|r|c|c|c|c|c|}
    \hline 
    & \multicolumn{5}{|c|}{$\mathcal{A}_{\rm PPM}$} \\ 
    & $s = 0.01$ & $s = 0.1$ & $s = 0.2$ & $s = 0.5$ & $s = 1$ \\ \hline
    IL1C \cite{b10} & 2.2250 & 2.0199 & \textbf{2.2215} & \textbf{4.1515} & \textbf{4.1766} \\
    Ensemble (\emph{us}) & \textbf{1.6982} & \textbf{1.8995} & 2.2643 & 4.4532 & 4.4647 \\ \hline
    \end{tabular}\\~\\~\\
    \begin{tabular}{|r|c|c|c|c|c|}
    \hline
    & \multicolumn{5}{|c|}{$\mathcal{A}_{\rm PUP}$} \\ 
    & $s = 0.01$ & $s = 0.1$ & $s = 0.2$ & $s = 0.5$ & $s = 1$ \\ \hline
    IL1C \cite{b10} & 0.3750 & 0.3543 & 0.3971 & \textbf{0.7186} & \textbf{0.7214} \\
    Ensemble (\emph{us}) & \textbf{0.2157} & \textbf{0.2414} & \textbf{0.2550} & 0.7421 & 0.8250 \\ \hline
    \end{tabular}

\end{table}

\section{Conclusions}\label{sec:conclusions}
In this work, we generalized the traditional experimental playground in which the notion of cross-correlation identity (CCI), applied to the estimation of TDOAs using blind channel deconvolution methods \cite{b10,b11,b14,b17,b18}, switching from the case $N = 2$ to $N > 2$. Our analysis shows that, by simply allowing for a increased number of microphones, the very same state-of-the-art method ILC1 \cite{b10} can be sharply boosted in performance (see Tab. \ref{tab:improvementsIL1C}) without requiring any change in the computational pipeline. 

We deem that our findings open up to a novel research trend in which CCI identities are better combined with the case $N > 2$, so that improvements in the error metrics can come from two different, yet complementary, factors: advances in the optimization standpoint and multiple CCI relationships. We warm-up the research efforts in this directions with two simple modifications of IL1C, showing that, with respect to an incremental addition of the microphones, the practitioners should preferred a late fusion ensemble mechanism - which has the understandable property of easing the peaks finding-based inference stage of \cite{b10}. 

%We presented a robust method in order to extend the IL1C algorithm for number of microphones $N>2$. We started solving the case $N=3$ using the standard IL1C algorithm and its incremental version. We then demonstrated that averaging the AIRs comings from the couples and using them to solve the case $N=3$ microphones, is a final robust tool to improve the metrics of the couples itself. Then, we've found a way to generalize the IL1C algorithm improving the performances increasing the number of microphones $N>2$ justifying this straightforward choice, testing it on synthetic and real signals. Next step is to increase the number of microphones and to use afterwords the method to reconstruct the geometry of the room through a real experimental set-up.

\vspace{12pt}
\color{red}
% IEEE conference templates contain guidance text for composing and formatting conference papers. Please ensure that all template text is removed from your conference paper prior to submission to the conference. Failure to remove the template text from your paper may result in your paper not being published.

\end{document}

%% file: defs.tex
\def\mat#1{\mathchoice{\mbox{\boldmath$\displaystyle\tt#1$}}
{\mbox{\boldmath$\textstyle\tt#1$}}
{\mbox{\boldmath$\scriptstyle\tt#1$}}
{\mbox{\boldmath$\scriptscriptstyle\tt#1$}}}

\def\vec#1{\mathchoice{\mbox{\boldmath  $\displaystyle\bf#1$}}
{\mbox{\boldmath  $\textstyle\bf#1$}}
{\mbox{\boldmath  $\scriptstyle\bf#1$}}
{\mbox{\boldmath  $\scriptscriptstyle\bf#1$}}}

\newlength{\colwidth}